 \documentclass[apj]{emulateapj}
\usepackage{color}
\usepackage{epsfig}
\usepackage{amsmath}
\usepackage{longtable}
\usepackage{graphicx}

\shorttitle{Relativistic Reconnection in a Proton-Electron Plasma}
\shortauthors{Guo et al.}

\doublespace
\begin{document}

\title{Efficient Production of High-energy Nonthermal Particles during Magnetic 
Reconnection in a Magnetically-dominated Ion-Electron Plasma}

\author{Fan Guo\altaffilmark{1}, Xiaocan Li\altaffilmark{1,2}, Hui Li\altaffilmark{1}, William Daughton\altaffilmark{1}, Bing Zhang\altaffilmark{3}, Nicole Lloyd-Ronning\altaffilmark{1}, Yi-Hsin Liu\altaffilmark{4}, Haocheng Zhang\altaffilmark{1}, Wei Deng\altaffilmark{1,3}}

\altaffiltext{1}{Los Alamos National Laboratory, Los Alamos, NM 87545, USA}

\altaffiltext{2}{Department of Space Sciences, University of Alabama in Huntsville, Huntsville, AL 35899, USA}

\altaffiltext{3}{Department of Physics and Astronomy, University of Nevada Las Vegas, Las Vegas, NV 89154, USA}

\altaffiltext{4}{NASA Goddard Space Flight Center, Greenbelt, MD 20771, USA}

\altaffiltext{5}{Astrophysical Institute, Department of Physics and Astronomy, Ohio University, Athens, OH 45701, USA}

\email{guofan.ustc@gmail.com}

\begin{abstract}
Magnetic reconnection is a leading mechanism 
for dissipating magnetic energy and accelerating nonthermal
particles in  Poynting-flux dominated flows.
In this letter, we investigate nonthermal particle acceleration
during magnetic reconnection in a magnetically-dominated 
ion-electron plasma using fully kinetic simulations. For an
ion-electron plasma with the total magnetization
$\sigma_0=B^2/(4\pi n(m_i+m_e)c^2)$, 
the magnetization for each species is 
$\sigma_i \sim \sigma_0$ and $\sigma_e \sim (m_i/m_e) \sigma_0$, respectively. We have studied the 
magnetically dominated regime by varying 
$\sigma_{e} = 10^3 - 10^5$ with initial ion and electron
temperatures $T_i = T_e = 5 - 20 m_ec^2$ and mass
ratio $m_i/m_e = 1 - 1836$. The results demonstrate that 
reconnection quickly establishes power-law energy distributions for 
both electrons and ions within several ($2-3$) light-crossing 
times. For the cases with periodic boundary conditions, the 
power-law index is $1<s<2$ for both electrons and ions. The hard spectra limit
the power-law energies for electrons and ions to be
$\gamma_{be} \sim \sigma_e$ and $\gamma_{bi} \sim \sigma_i$, 
respectively. The main acceleration mechanism is a Fermi-like
acceleration through the drift motions of charged particles.
When comparing the spectra for electrons and ions in momentum space, the spectral indices $s_p$ are identical
as predicted in Fermi acceleration. We also find that the bulk flow can carry
a significant amount of energy during the simulations.
We discuss the implication of this study 
in the context of Poynting-flux dominated jets and pulsar 
winds especially the applications for explaining the nonthermal 
high-energy emissions.

\end{abstract}


\section{Introduction}

It is thought that magnetic field is of 
central importance in pulsar wind nebulae (PWNe) and jets from 
active galactic nuclei (AGNs) and gamma-ray bursts (GRBs). 
The initially
launched relativistic flows are Poynting-flux dominated 
\citep{Spruit2010}. Meanwhile, these
astrophysical phenomena are observed to have 
non-thermal spectra that arise from various
emission processes of energetic particles. 
For example, leptonic modeling of TeV blazar emissions 
requires a power-law electron distribution with 
energy $\gamma_e\sim10^3-10^5$ 
\citep{Bottcher2013}.
The spectra of gamma-ray bursts during the
prompt phase suggest that the 
radiating particles have Lorentz factors 
$\gamma_e\sim10^3-10^7$ \citep{Daigne1998,Uhm2014}. 
In hadronic models, protons also need to be accelerated
to high energies and may be related to ultra-high-energy 
cosmic rays \citep{Bottcher2013}.
Understanding how the energy stored in 
magnetic fields being converted into 
nonthermal particle energy is essential to 
interpret observed emissions in those 
high-energy astrophysical processes. 

Recent studies of GRBs 
\citep{Zhang2011,McKinney2012,Kumar2015}, AGNs 
\citep{Giannios2009,Zhang2015a} and 
PWNe \citep{Uzdensky2014} have shown 
that models that are magnetically dominant can 
explain much of the observed data. It is
estimated that the magnetization parameter
$\sigma\equiv B^2/(4\pi w)$
can be large and the 
Alfv$\acute{e}$n speed approaches the light speed $v_A\sim c$. Here $B$ is the magnitude of magnetic field
and $w$ is the enthalpy. To explain the observed 
high-energy emissions, an efficient conversion
from energies in the magnetized flow 
into nonthermal particles is required 
\citep[e.g.,][]{Zhang2011}. 
Collisionless shocks that are considered to be
efficient for particle acceleration in 
weakly magnetized plasmas are inefficient 
in dissipating energy and accelerating nonthermal 
particles in magnetically-dominated flows
\citep{Sironi2015,Zhang2015b}. 

Magnetic reconnection is a major candidate that 
unleashes magnetic energy in
magnetized flows through 
reorganizing magnetic topology 
\citep{Zweibel2009}. In the 
relativistic limit, the magnetic energy gets
strongly dissipated in the reconnection region 
\citep{Blackman1994,Lyutikov2003,Comisso2014,Liu2015}. 

Theory of nonthermal particle acceleration in reconnection 
layers is still evolving
\citep{Pino2005,Drake2006,Drury2012,Guo2014,Zank2014}. 
Much progress has been made through particle-in-cell (PIC)
kinetic simulations that self-consistently
model the microscopic physics of collisionless reconnection
and nonthermal plasma energization. Earlier studies focus on the low-$\sigma$ regime 
($\sigma<1$), and identified several basic
acceleration mechanisms such as direct 
acceleration at X-lines
\citep{Fu2006,Huang2010}, Fermi-type acceleration 
associated with plasmoids
\citep{Drake2006}, and further acceleration in 
plasmoid coalescence regions \citep{Oka2010}. 
Simulations with mildly relativistic parameters
($\sigma\sim1$) have uncovered 
particle acceleration with power-law energy 
spectra in X-lines
\citep{Zenitani2001,Bessho2012}, but no significant
power-law distribution when the 
spectra are integrated over the whole domain \citep{Zenitani2007,Liu2011,Kagan2013}. Recently, there is an explosion of studies on relativistic magnetic 
reconnection in the magnetically-dominated regime ($\sigma\gg1$) \citep{Cerutti2012,Sironi2014,Guo2014,Guo2015,Melzani2014b,Werner2014,Liu2015}, primarily motivated by the discovery of the spectacular Crab flares \citep{Abdo2011,Tavani2011} and development of the magnetically-dominated models.
Several simulations have reported hard power-law
distributions with spectral index $1\leq s\leq2$ when $\sigma\gg1$
\citep{Sironi2014,Guo2014,Guo2015,Melzani2014b,Werner2014} 
\footnote{Several papers reported $s > 2$ but the spectra are 
plotted as a function of Lorentz factor $\gamma$ rather than 
kinetic energy $\gamma-1$. When plotting energy spectrum as a function of 
$\gamma-1$, we did not find obvious power-law distribution with $s>2$ for 
simulations with periodic boundary conditions.}. 
These new simulations claim power-law distributions in the whole 
reconnection region, suggesting reconnection in the magnetically-dominated regime
may be a strong source of nonthermally particles. 
\citet{Guo2014,Guo2015} used a
force-free current sheet that does not require 
the hot plasma population in a Harris sheet and showed that the
energy spectra of particles within the entire
reconnection layer resembles a power law. Through calculating the guiding-center drift motions of
all the simulated particles, we have demonstrated that
the primary acceleration mechanism is
a first-order Fermi mechanism resulting from the particle
curvature drift motion along the direction of
the $-\textbf{v}\times \textbf{B}/c$ electric field induced by the relativistic
flows. The acceleration leads to the formation 
of hard power-law spectra 
$f\propto(\gamma-1)^{-s}$ with $s\sim1$ for sufficiently strong magnetic 
field and large simulation domains.
We have developed an analytical model to describe 
the simulation results and derived a general condition for the formation of the
power-law distributions, i.e., the acceleration
time scale needs to be shorter than the time scale
for particles injected into the reconnection region $\tau_{acc}<\tau_{inj}$. 
The solution may also explain simulations 
from the Harris current layer
\citep{Sironi2014,Melzani2014b,Werner2014}. 
The power-law distribution
has also been found in nonrelativistic
reconnection simulations with a 
low-$\beta$ ($\beta=8\pi nkT/B^2\ll1$) plasma \citep{Li2015}, 
showing the mechanism can be applied in a larger parameter range.

Most PIC simulations of relativistic magnetic
reconnection focused on pair plasmas. In general 
the plasma composition is uncertain and can be either 
electron-positron pairs or electron-ion plasmas. 
While proton-electron reconnection has been
studied by \citet{Melzani2014a,Melzani2014b} with a limited
parameter range ($m_i/m_e=1-50$, $\sigma_0=B^2/4\pi(m_i+m_e)nc^2=1-10$), 
here we show results from two-dimensional PIC simulations for a magnetically
dominated ion-electron plasma, the
magnetization for each species is $\sigma_i\sim\sigma_0$ and
$\sigma_e\sim(m_i/m_e)\sigma_0$. We study the 
magnetically dominated regime for a range of $\sigma_0$ with different ion 
and electron temperatures and mass ratio $m_i/m_e$
up to $1836$. We demonstrate that reconnection quickly
establishes power-law distributions for both electrons and 
ions within several light-crossing times. The power-law index is $1<s<2$ for 
both electrons and ions in the cases with periodic boundaries. 
The break energies for electrons and ions are
$\gamma_{be}\sim\sigma_e$ and $\gamma_{bi}\sim\sigma_i$.
For the antiparallel case we study, the main acceleration 
mechanism for particles in the power-law is a Fermi-like acceleration through drift 
motions of charged particles, consistent with earlier work
\citep{Guo2014,Guo2015}. 
The numerical methods are presented in Section 2. 
Section 3 discusses the main results 
of the paper. In Section
4, we discuss the implications of the study in the context of high-energy
astrophysics.  Our conclusions are summarized in Section 5.

\section{Numerical Methods}
The simulations start from a magnetically-dominated force-free current sheet. 
Previous studies have shown that the intense current sheets can
develop through various processes such as striped wind geometry, field-line 
foot-point motion, and turbulence cascade
\citep{Coroniti1990,Titov2003,Makwana2015}.
The initial condition of our simulations is a force-free current layer with 
$\textbf{B}=B_0\text{tanh}(z/\lambda)\hat{x}+B_0\text{sech}(z/\lambda)\hat{y}$\citep{Guo2014,Guo2015}. This corresponds to a rotating 
magnetic field with a $180^\circ$ change in direction within a thickness of $2 \lambda$. The initial distributions are spatially uniform with density $n_0$ and relativistic Maxwellian in energy space. The current density $\textbf{J}=en_0(\textbf{U}_i-\textbf{U}_e)$ is represented
by a drift velocity $\textbf{U}_i=-\textbf{U}_e$ that satisfies Ampere's law.

The simulations assume an electron-ion plasma with 
mass ratio $m_i/m_e=1\rightarrow1836$ and we mainly focus
on the case with $m_i/m_e\gg1$. For simplicity, the initial thermal temperatures for ions and electrons are assumed to be the same and varied in a
range $T_i=T_e=5\rightarrow20m_ec^2$ to examine its effect on the resulting energy spectra. This temperature is similar to the temperature inferred for AGN jets \citep{Homan2006}. The full particle simulations are performed using the VPIC code
\citep{Bowers2009}, which explicitly solves Maxwell equations and the relativistic equation of motion for the particles. In the simulations, $\sigma_e$ is adjusted 
by changing the ratio of the nonrelativistic electron gyrofrequency $\Omega_{ce0}=eB/(m_ec)$ 
to the nonrelativistic electron plasma frequency $\omega_{pe0}=\sqrt{4\pi ne^2/m_e}$, 
$\sigma_e\equiv B^2/(4\pi n_em_ec^2)=(\Omega_{ce0}/
\omega_{pe0})^2$ and the total magnetization is $\sigma_0\sim(m_e/m_i)\sigma_e$. 
We primarily discuss simulations with $\sigma_e=10^3\rightarrow10^5$. The 
cell sizes are chosen so $\Delta x=\Delta z<0.1d_e$. 
Here $d_e=\sqrt{\gamma_0}d_{e0}$, in which 
$\gamma_0=1+3T_e/2m_ec^2$ and $d_{e0}=c/\omega_{pe0}$.
All the simulations have $100$ electron-ion pairs in each 
cell. Similar to numerous earlier studies, the boundary conditions in the x-direction are periodic for both fields and particles, and in the $z$-direction the boundaries are conductive 
for the field and reflect particles when they reach the boundaries.
A long-wavelength perturbation
with $B_z=0.05B_0$ is included to 
initiate reconnection. 
To ensure numerical convergence, 
it is essential to have enough particles per cell and spatial grids to avoid
excessive numerical heating \citep{Guo2015}. We have monitored the 
numerical energy conservation to ensure that effect
does not influence our results. A more comprehensive study will be reported 
in another publication (Guo et al. 2016 in preparation).

\section{Simulation results}

Figure $1$ shows the current 
layer evolution in the simulation with 
$\sigma_e=10^4$, $T_i=T_e=10m_ec^2$, and $m_i/m_e=100$ so $\sigma_0\sim100$. The domain size 
is $L_x\times L_z=4000d_{e0}\times2000d_{e0}$. This is the representative run that we will discuss in detail. The left panels and right panels show the current density and velocity in the x-direction at $\omega_{pe0}t=2000$, $3000$, and $5000$, respectively. The current sheet thins down under the influence of the perturbation and then breaks into many fast-moving secondary plasmoids due to the growth of secondary tearing instability. The thinning phase lasts until $\omega_{pe0}t\sim2000$ and rest of the simulation is dominated by the evolution of plasmoids. These plasmoids coalesce and merge into a single island on the order of the system size. 
The whole process is similar to the case with pair plasmas \citep{Daughton2007} as long 
as the simulation domain is sufficiently large to contain several secondary tearing modes. The simulation also shows the development of fast bulk plasma flows with bulk Lorentz factors $\Gamma=1/\sqrt{1-(V/c)^2}$, associated with plasmoids \citep{Liu2015,Guo2015}. As we will discuss
below, the fast moving plasmoids contain most of the accelerated energetic 
particles, which likely generates high-energy nonthermal emissions with strong 
beaming effects. We also observed magnetosonic waves generated as the 
plasmoids interact with each other, similar to recent MHD simulations 
\citep{Yang2015}.

\begin{figure*}
\begin{center}
\includegraphics[width=0.9\textwidth]{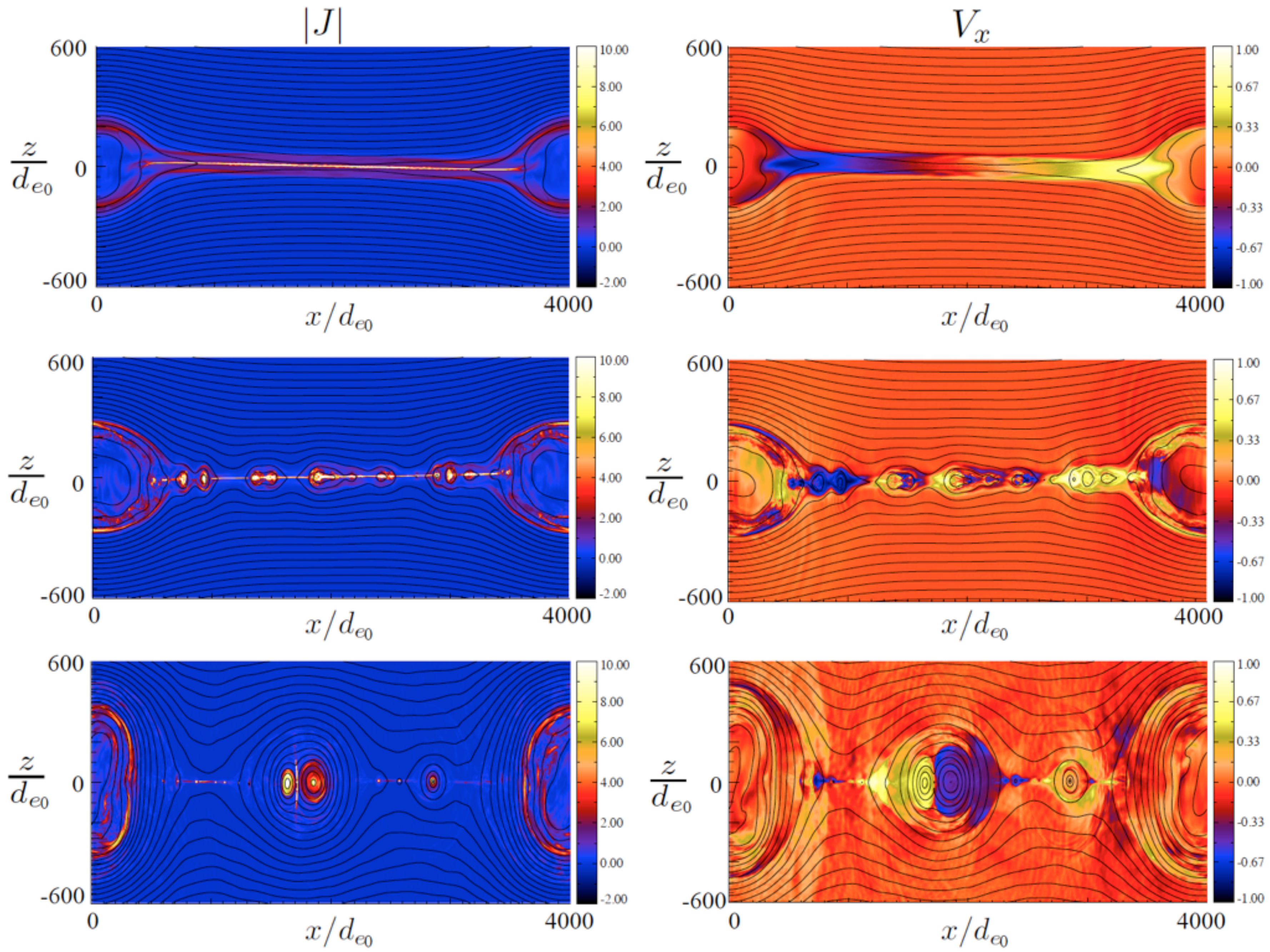}
\caption{Time evolution of 2D current density and bulk flow velocity along the x-direction for the case $\sigma_0=100$, $m_i/m_e=100$ at $\omega_{pe}t=2000,\,3000$, and $5000$.  The secondary tearing mode breaks the elongated current sheet and produce multiple plasmoids with the associated maximum bulk flow gamma of a few.}
\end{center}
\end{figure*}

Figure 2 summarize the plasma energetics for the representative run. In 
Figure 2(a), the magnetic energy in the reconnecting component drops more
than $60\%$ within $2-3$ light crossing times and most of it gets 
converted into plasma energies. Ions gain slightly more energy than 
electrons $E_{ki}/E_{ke}\sim1.1$. Increasing $m_i/m_e$ to $1836$, we find 
$E_{ki}/E_{ke}\sim1.5$, meaning in this case the energy partition does not strongly 
depend on the ion-to-electron mass ratio. Figure 2(b) 
shows the evolution of bulk energy, internal energy, and total kinetic energy. 
The internal energy is defined as the kinetic energy
in the local plasma frame. The bulk energy is energy associated with the 
plasma bulk motion. The figure shows that the bulk flow can carry a nontrivial
portion of the total plasma energy during the energy conversion. Initially it exceeds
the internal energy and still 
contains $5-10\%$ of the total kinetic energy toward the end of the simulation. 

Figure 3 presents the energy spectra for both electrons and ions at 
different simulation times for the 
representative case. In energy space, electron distribution 
shows a power-law distribution with $s\sim1.35$, whereas the ion distribution 
shows a power law with another slope $s\sim1.2$ starting from several times of
the initial thermal energy. The break energies correspond 
to $\gamma_{be}\sim\sigma_e$ and $\gamma_{bi}\sim\sigma_i$ and above that
the spectra resemble an exponential cutoff.  During the initial phase ($\omega_{pe0}t<3000$), the flat spectrum is associated with the acceleration
by the parallel electric field during the thinning process and later the 
acceleration is dominated by Fermi acceleration associated with plasmoid motion 
(see discussion below). Figure 3(c) shows the momentum distribution $f(p)p^2$ for 
both of the species. We find that if the
two distributions are plotted in momentum space, they 
give almost the same slope 
$f(p) \propto p^{-s_p}$ and $s_p\sim3.35$. This is a general feature 
of Fermi acceleration but not obviously 
predicted by other mechanism proposed to explain power-laws generated 
in reconnection regions \citep{Sironi2014,Werner2014}.

\begin{figure*}
\includegraphics[width=\textwidth]{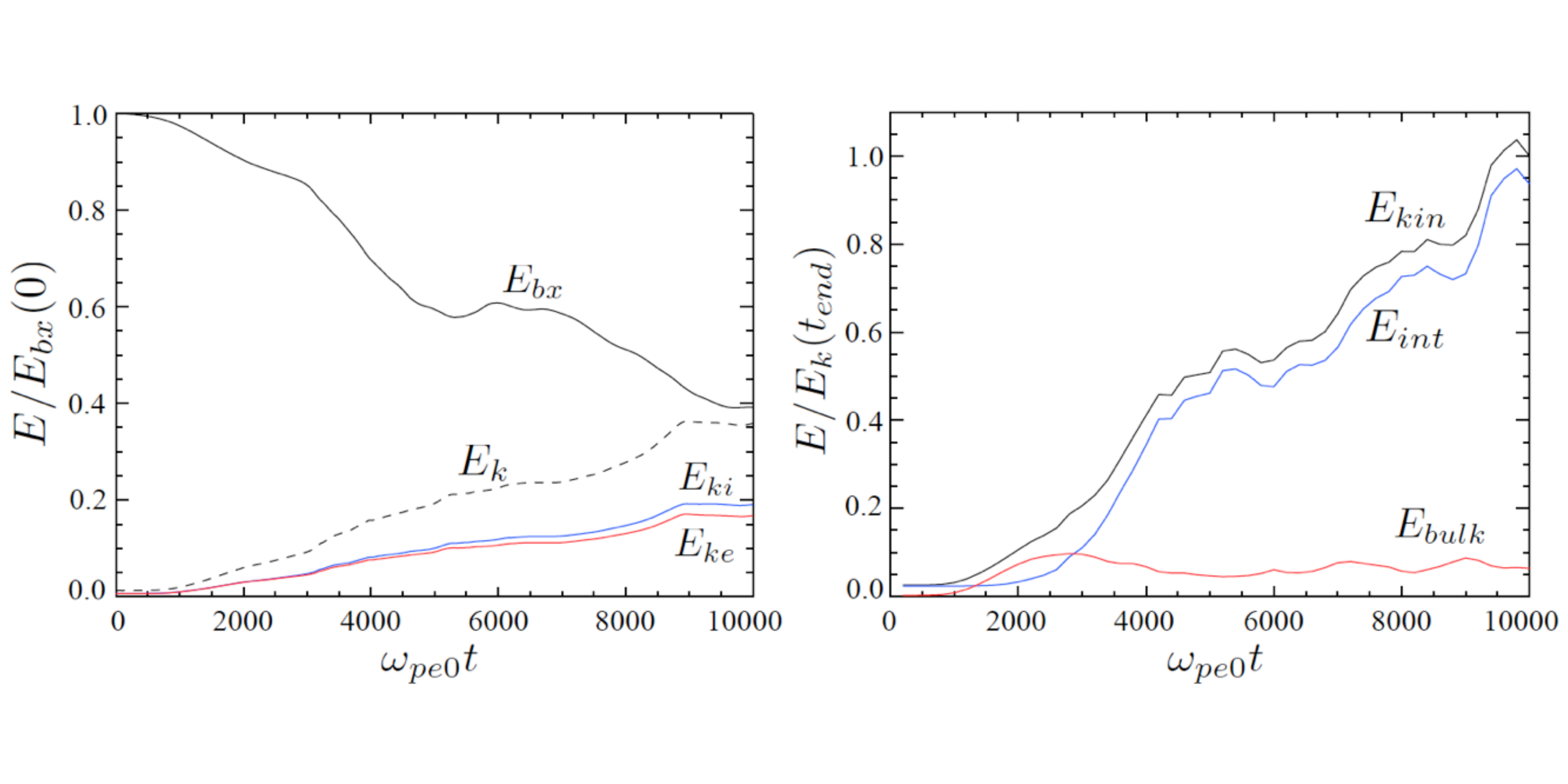}
\caption{Energetics for the run with $\sigma_0=100$, $m_i/m_e=100$. Left panel: evolution of reconnecting magnetic energy $B_x^2/8\pi$, ion kinetic
energy and electron kinetic energy. Right panel: the evolution of total kinetic energy, internal energy, and bulk flow energy.}
\end{figure*}

Figure 4 compares electron energy spectra for simulations with
different (a) $\sigma_e$, (b) initial temperature, 
(c) box size, and (d) mass ratio. Results show that for a sufficiently large domain,
both electrons and ions get efficient accelerated and form a 
power-law distribution. For different $\sigma_e$, we observe slightly harder
spectrum as the available magnetic energy increases compare to the 
plasma energy. The trend is similar as we reduce the thermal 
energy. Panel (c) shows that for a larger simulation domain, the power-law 
distribution becomes fully developed. As the box size increased to $L_x=4000d_{e0}$ for $m_i/m_e=100$, 
the energy spectra are similar for larger systems.
We also find that as long as the simulation domain is sufficiently large, 
the mass ratio does not significantly influence the electron spectrum.
As for the spectral index, we find that the simulations give $1<s<2$ for 
the periodic closed system typically used for reconnection studies. This,
however, may be changed for the case with open boundary conditions 
\citep[][]{Daughton2006} where the particle escape from the boundaries 
could lead to a softer spectrum \citep{Guo2014,Guo2015}.

\begin{figure}
\includegraphics[width=0.5\textwidth]{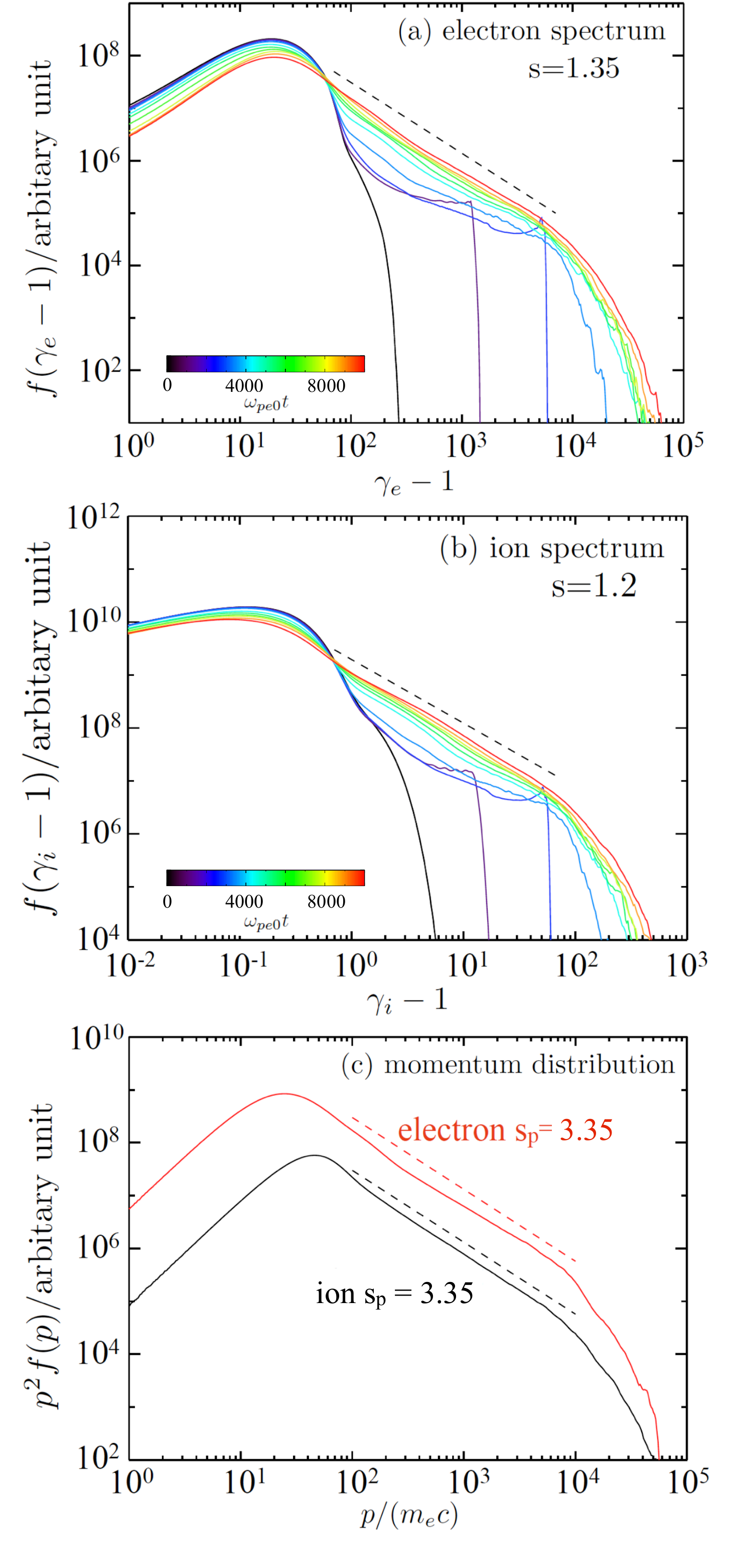}
\caption{Energy distributions $f(E)$ and momentum distributions $f(p)p^2$ for both ions and electrons for the representative run with $\sigma_0=100$ and $m_i/m_e=100$.}
\end{figure}

\begin{figure*}
\includegraphics[width=\textwidth]{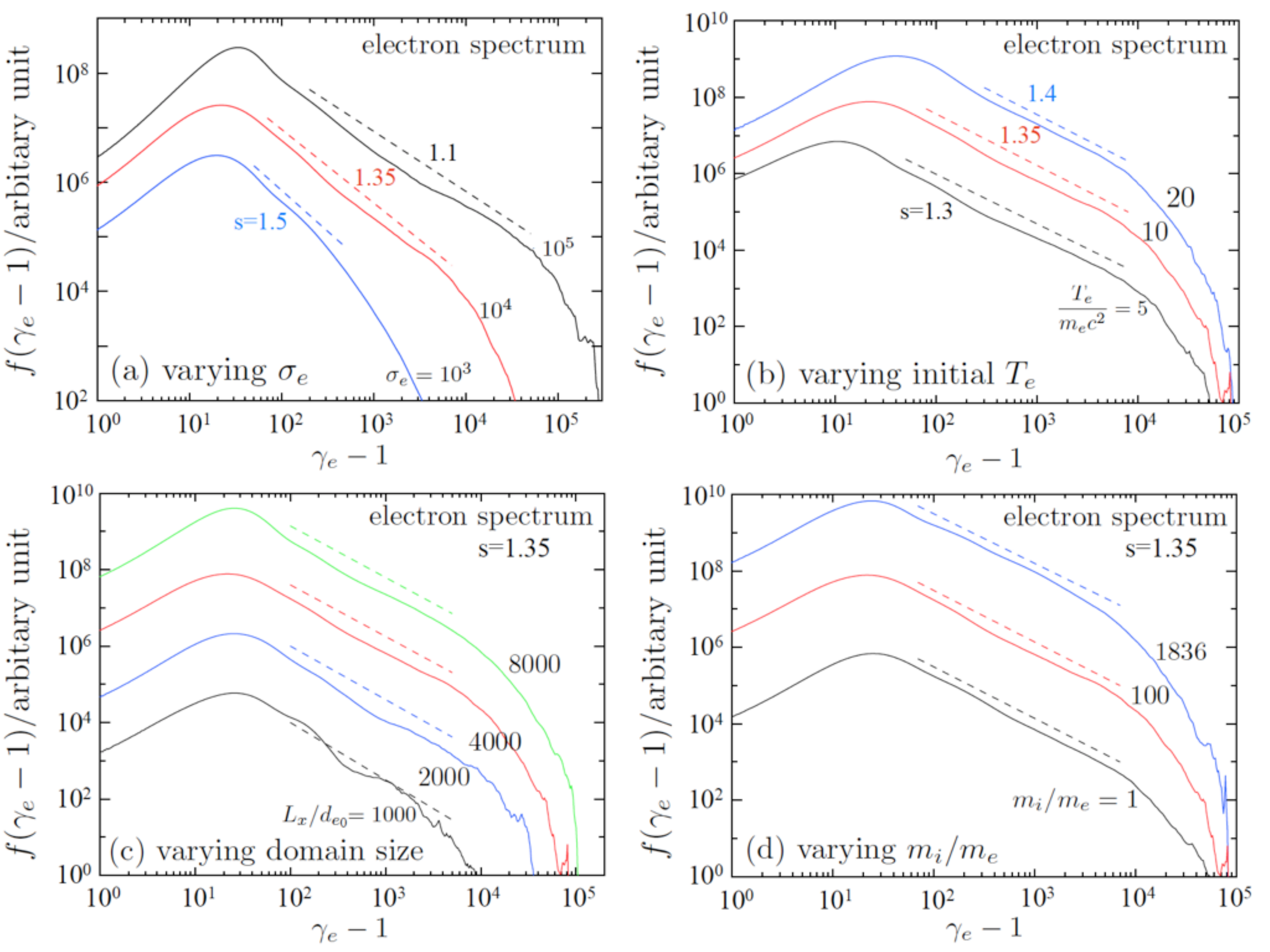}
\caption{Electron energy spectra for cases with different $\sigma_e$, different temperature, different box size, different mass ratio}
\end{figure*}

Now we briefly discuss the acceleration of protons 
and electrons using a 
particle tracking module newly implemented 
in VPIC. We find that for particles 
initially in the current sheet, they mainly 
experience strong direct acceleration through 
parallel electric field. Typical examples for both electrons
and ions are shown in Figure 5(a). For particles that get 
accelerated to the highest energies $\gamma_{max}$ in the 
simulation, this is the main type of 
acceleration. However, for the accelerated 
particles that form the power-law distribution 
$\gamma < \gamma_b$,
the main acceleration pattern is a Fermi-like 
process where particles gain energy by bouncing
back and forth a few times associated with the 
motion of plasmoids as the plasmoids evolve and coalesce. 
In Figure 5(b) we show typical particle 
trajectories that resemble the Fermi mechanism. 
The acceleration process for both electrons and ions are 
very similar. We verified that the size 
of the largest island is well above the gyroradius of 
power-law ions with the real mass ratio.
In a forthcoming paper, we will discuss the statistics
of particle acceleration in detail similar to previous
studies \citep{Guo2014,Li2015}.

\begin{figure*}
\includegraphics[width=\textwidth]{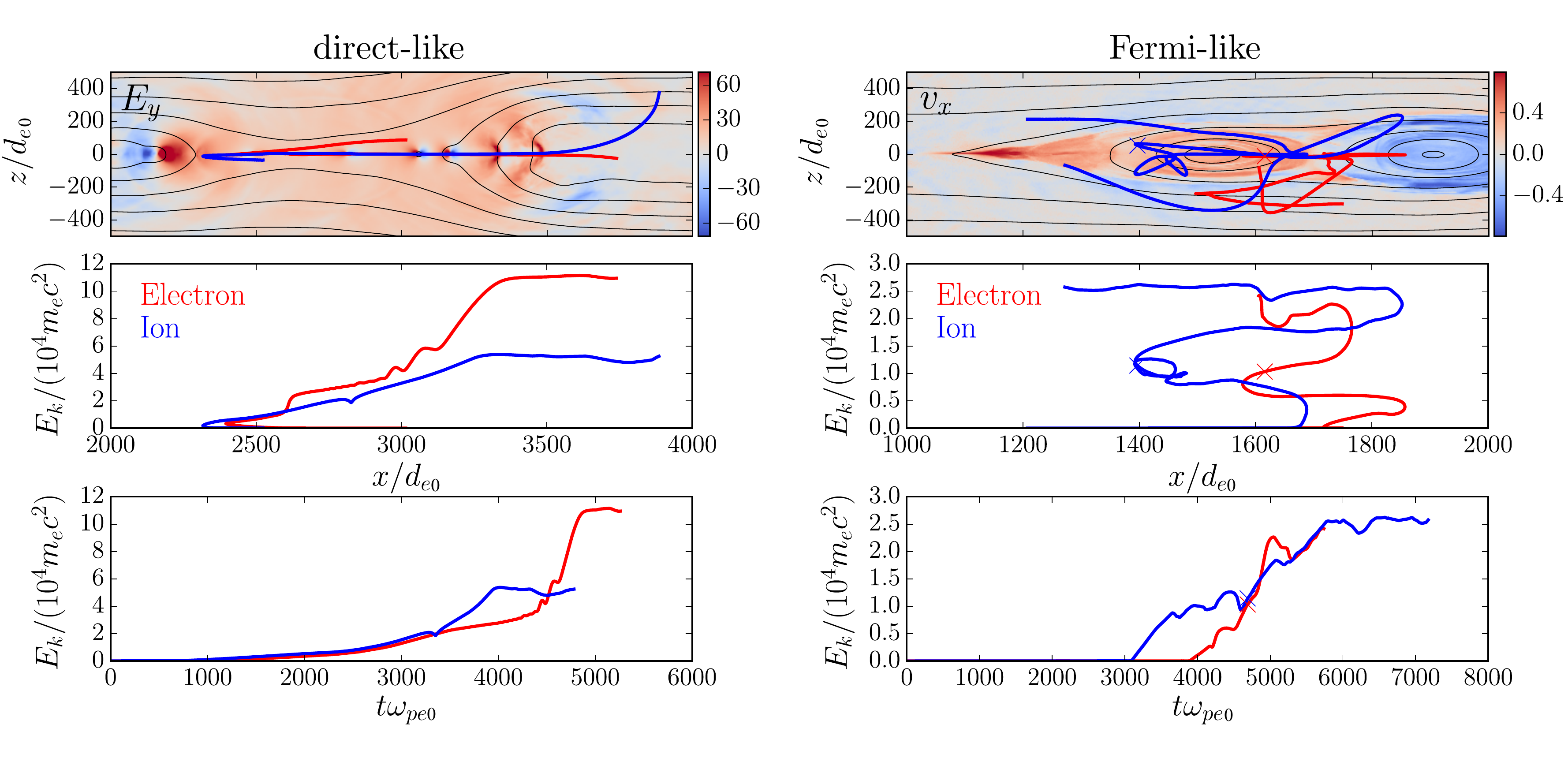}
\caption{Typical trajectories for accelerated particles. Left panel shows the most accelerated particles whose trajectories similar to the direct acceleration. The color-coded contour shows the electric field in the $y$-direction normalized using $m_ec\omega_{pe}/e$. Right panel shows the ones typical for power-law energy range similar to the Fermi acceleration. The open circles indicate the starting points and the cross signs show the particle locations at the same time step as the field contour.}
\end{figure*}

\section{Discussion and implications}

It is thought that as relativistic jets or pulsar winds are
launched, initially the energy flow are magnetically 
dominated and then the magnetic energy gets converted
into plasma energy \citep{Spruit2010}.
During this process strong sheared current sheets may form \citep{McKinney2012} and magnetic energy gets
converted into plasma energy through magnetic reconnection
and produces nonthermal emission.
Our results imply that the observed nonthermal emission 
from AGNs, GRBs, and PWNe could come from strong
particle acceleration in magnetic reconnection regions. 
For example, when modeling the emissions from TeV blazars, 
it is required that the electron spectrum is a power-law
distribution with energy range 
$\gamma_e=10^3-10^5$ \citep{Bottcher2013}. Recent observations 
show that strong multiwavelength flares are sometimes 
accompanied by strong polarization variations, indicating the active 
participation of magnetic field during flares \citep{Zhang2015a}. 
Our calculation
shows the development of power-law distribution with break
energy $\gamma_{be}\sim\sigma_e$ regardless of the ion-to-electron 
mass ratio, suggesting that the nonthermal electrons
can be produced when $1\lesssim\sigma_0\lesssim100$ ($1800\lesssim\sigma_e \lesssim 1.8\times10^5$) for real mass 
ratio. Our simulations show the spectral index is 
$1<s<2$, which is sometimes reported in TeV blazars 
\citep[][]{Hayashida2015,Yan2015}. 
Most of energetic particles 
are contained in plasmoids and the bulk flow Lorentz factor $\Gamma$ can 
be a few so we expect radiation from the nonthermal particles
has strong beaming effects that may help explain fast 
variability observed in TeV blazars
\citep{Giannios2009}.

In GRBs, the spectra of prompt emission suggest that the 
radiating electrons can reach $\gamma_e\sim 100$ and form
a non-thermal energy distribution. In our simulation, the power-law distribution 
starts from several times of the thermal energy ($\gamma_{1}\sim50-200$) and 
extends to high energies ($\gamma_{2}\sim\sigma_e$) up to $10^5$. As the small-scale reconnection events continuously occur, electrons can 
be continually accelerated so the so-called ``slow cooling'' regime is likely to be valid. The highly efficient acceleration may 
help to solve the efficiency problem in GRB models and 
the relativistic 
bulk motion may explain fast variability observed in some GRB events
\citep{Zhang2011,Deng2015}.

Theoretical analysis has shown that strong particle acceleration 
and/or beaming effect in the reconnection region
can explain the gamma-ray Crab flares \citep{Abdo2011,Tavani2011}. 
Our simulation shows that reconnection is a credible process 
that drives both strong particle acceleration and relativistic bulk motion.
The strong energy conversion from magnetic energy into the plasma kinetic 
energy during relativistic magnetic reconnection may also 
help understand the so-called $\sigma$-problem, where the energy conversion 
is needed for dissipating the Poynting-flux dominated flow 
\citep[e.g.,][]{Coroniti1990}.

In some hadronic models, a proton spectrum with
$s<2$ may be required to model TeV blazars \citep{Bottcher2013}. Our results show that ion energy gain is comparable to the electron energy gain and most of the energy goes into a nonthermal power-law spectrum. The maximum
energy of protons $\gamma_{imax}$ 
depends on the system size and the
largest gyroradius can be comparable to the system size.
This suggests that magnetic reconnection can be
a viable mechanism for providing energetic protons in
hadronic models.

Most observed energy 
spectra are considerably softer than the simulated results.
Moreover, the hard power-laws ($s<2$) obtained in this and other works 
are only in a limited energy range. While the maximum
energy of particles grows with time, the power-law break energy does 
not and is limited by the 
amount of dissipated magnetic energy 
$\gamma_{be}\sim(m_i/m_e)\gamma_{bi}\sim[\delta\sigma_e(2-s)]^{1/(2-s)}$, where 
$\delta\sim0.2$ is the fraction of the dissipated magnetic energy channeled into 
each species. This poses a serious problem that 
has to be addressed in the future. 
In our earlier papers \citep{Guo2014,Guo2015}, we have 
shown that particle escape from the acceleration 
region could lead to a softer spectrum. Since our current 
simulations employ periodic systems, the acceleration
tends to give the hardest spectrum. This suggests more 
realistic open boundary conditions \citep[e.g.,][]{Daughton2006}
that include the effect of
particle escape from the simulation domain is necessary.
Also, our results are tested for a domain 
size up to $L\sim 10^4d_{e0}$, but for larger system the spectrum
could still evolve as acceleration continues.



\section{Summary}

Astrophysical reconnection sites have long been expected to be a strong source of nonthermal particles. Our current study has demonstrated that relativistic magnetic reconnection is efficient at generating nonthermal
particles though the underlying energy conversion and
acceleration process in a magnetically dominated proton-
electron plasma. During reconnection, the energy distributions for both ions
and electrons develop power-law spectra, which is important to understand
the nonthermal emission from objects like PWNe, AGNs,
and GRBs. We also find that during the active reconnection phase the outflow 
speeds approach the light speed and the bulk Lorentz factor can reach a few, which may help explain the observed fast variability and high luminosity
emissions in several systems. These findings
on plasma energization and particle acceleration in reconnection regions further demonstrate the important effect of
magnetic reconnection in high-energy astrophysical systems.


\section*{Acknowledgement}
We are grateful to Jonathan Jara-Almonte, who developed an 
initial version of the particle tracking module. 
This work is supported by 
the DOE through the LDRD program at LANL and DOE/OFES support 
to LANL in collaboration with CMSO. X.L is supported by NASA Headquarters under the NASA Earth
and Space Science Fellowship Program-Grant NNX13AM30H. Contributions from W.D. were supported by NASA from the Heliophysics Theory Program.
N.L-R. is supported by the M. Hildred 
Blewett Fellowship of the APS. Simulation resources are provided by LANL institutional
computing.


\end{document}